# Full characterization of spin-orbit coupled photons via spatial-Stokes measurement

Bing-Shi Yu,[1,*] Hai-Jun Wu,[1,*] Hao-Ran Yang,[1] Wei Gao,[1] Carmelo Rosales-Guzmán,[1] Bao-Sen Shi,[1,2] and Zhi-Han Zhu[1,†]

[1] *Wang Da-Heng Collaborative Innovation Center, Heilongjiang Provincial Key Laboratory of Quantum manipulation & Control, Harbin University of Science and Technology, Harbin 150080, China*
[2] *CAS Key Laboratory of Quantum Information, University of Science and Technology of China, Hefei, 230026, China*

Characterization and analysis of spin-orbit coupled (SOC) states, as a measurement problem, play a vital role in research on the modern optics and photonics based on structured light. Here, we demonstrate determination of photonic SOC states via spatial-Stokes measurement, in which two spatial complex probability amplitudes of spin-dependent spatial modes within SOC states and their relative (intramode) phase can be measured directly. Compared with the standard quantum-state tomography, by avoiding wavefront-flattening operations, the apparatus can completely record photons' realistic SOC structure, leading to a more accurate and precise determination of wavefunction. This simple and general approach for SOC state determination can provide a powerful toolkit for in-situ measuring photonic SOC state, characterizing the quality of SOC light source and associated geometric-phase devices.

The term photonic spin-orbit coupling (SOC) refers to the electromagnetic interaction at subwavelength scales in which the two components of photonic angular momenta — i.e., spin angular momenta (SAM) and orbital angular momenta (OAM) — couple with each other forming an entangled structure [1-3]. This intrinsic entanglement within photons indicates an inhomogeneous structure inside photons that features a spatial-variant state of polarization (SOP) and, therefore, is broadly termed as vector (polarization) modes or structured light [4-6]. Recently, this long neglected inhomogeneity has brought a paradigm shift in the optical community, leading to the realization of myriad of optical phenomena and functionalities by exploiting photonic SOC, such as, observing subwavelength-scale microstructures, enhancing the capabilities of optical communication and metrology, shaping and controlling the structure and propagation of light, and promoting the research in fundamental physics [7-24]. In this emergent field of modern optics, characterization and analysis of vector modes play a vital role for its importance in both fundamental and practical aspects.

To fully characterize the SOC structure of a vector mode, one need to determine its wavefunction, i.e., SOC state. The standard approach for this task is the well-established Quantum state tomography (QST), in which one has to use a series of real-valued observables, i.e., mutually unbiased (MU) observables, to determine the complex-valued wavefunction indirectly. For SOC states, the MU-observables are called high-order Stokes parameters (HOSPs), and this specific QST is accordingly termed high-order Stokes tomography (HOST) [20]. To realize HOST experimentally, to date, the most feasible way is to employ q-plates or equivalent devices to perform a series of spin-dependent and phase-locked OAM detections [10-12]. So that the intramode phase of wavefunctions can been coupled to the apparatus and transferred to outcomes. However, due to the bottleneck of OAM detection based on wavefront-flattening measurements [25,26], HOST can only apply to some simply predefined states and its accuracy and precision are also restricted.

An alternative approach for wavefunction determination is a quantum state direct measurement, where, by sequential measurement for two complementary observables, the apparatus can directly record the complex probability amplitudes of wavefunction. Therefore, it has been recognized as a promising way for *in situ* characterization of quantum states and dynamic processes [27-32]. Particularly, a series of direct measurement for non-entangled photonic states have been reported recently [27-29]. Among them, the first observation has been done in the context of 'weak measurement', where the system is weakly coupled (or entangled) with an incompatible observable or degree-of-freedom, providing, in this way, an interface of complementary observation for the following measurement. In this work, we demonstrate direct measurement of photonic SOC states via spatial-Stokes measurement (SSM), where the SOC photons are observed in three MU-bases of the spin subspace. The results show that, due to avoiding wavefront-flattening operations in experiment, the apparatus can record two spatial complex probability amplitudes of the orthogonal spatial modes within a vector mode and their intramode phase directly, or equivalently to acquire a realistic vector profile consisting of both intensity and polarization profiles, leading to high precision and accuracy in SOC state determination and light-field characterization.

*Theory.* — Here, to demonstrate the principle in a simple way and without losing generality, we choose well-known cylindrically symmetric vector (CV) modes as target states to be measured. Their full wavefunctions can be expressed as

$$|\psi_{CV}(r,\theta,z,\ell)\rangle = a_1|\hat{\mathbf{e}}_R, \psi_R^\ell\rangle + a_2|\hat{\mathbf{e}}_L, \psi_L^\ell\rangle \\ = A_{|\ell|}(r,\theta,z)\left[a_1|\hat{\mathbf{e}}_R, +\ell\rangle + a_2|\hat{\mathbf{e}}_L, -\ell\rangle\right], \quad (1)$$

where $a_1$ and $a_2$ ($a_1^2 + a_2^2 = 1$) are two complex probability amplitudes of SOC states; $\hat{\mathbf{e}}_{L,R}$ denote the left- and right-hand circular polarizations carrying a SAM of $\pm\hbar$ per photon; and $\psi_{R,L}^\ell = A_{|\ell|}(r,\theta,z)e^{i(kz\pm\ell\theta)}$ are corresponding spin-dependent spatial modes carrying an amount $\pm\ell\hbar$ of OAM per photon. Note that $\psi_{R,L}^\ell$ carry opposite topological charges ($\pm\ell$) but have the same spatial amplitude $A_{|\ell|}(r,\theta,z)$. Moreover, the vector profiles (intensity and polarization) of CV modes are rotationally invariant for $\ell = +1$, therefore, it is of great relevance for alignment-free quantum communication [10,11]. If we do not concern about $A_{|\ell|}(r,\theta,z)$, the concise SOC state and associated density matrix of a CV mode can be expressed as

$$|\psi_{SO}\rangle = a_1|0\rangle + a_2|1\rangle \text{ and } \hat{\rho}_{SO} = |\psi_{SO}\rangle\langle\psi_{SO}|, \quad (2)$$

respectively, where $|0\rangle = |\hat{e}_R, +\ell\rangle$ and $|1\rangle = |\hat{e}_L, -\ell\rangle$ are a group of MU-bases of the two-level SOC space, and the other two are $(|0\rangle \pm |1\rangle)/\sqrt{2}$ and $(|0\rangle \pm i|1\rangle)/\sqrt{2}$. This bipartite system can be described visually by states on the higher-order Poincaré sphere (HOPS) [20], which is the topological equivalent of the Poincaré sphere and associated Stokes parameters (SPs) for polarization. More specifically, $|0\rangle$ and $|1\rangle$ are the two poles of the HOPS, with the expectation values of $\langle\psi_{SO}|\hat{\sigma}_{0-3}^{SO}|\psi_{SO}\rangle$, or $\text{tr}(\hat{\rho}_{SO}\hat{\sigma}_{0-3}^{SO})$, specifying the location of a state on the sphere called HOSPs, which are given by

$$\begin{cases} S_0^\ell = |\langle\hat{e}_R, +\ell|\psi_{SO}\rangle|^2 + |\langle\hat{e}_L, -\ell|\psi_{SO}\rangle|^2 \\ S_1^\ell = 2\text{Re}(\langle\hat{e}_R, +\ell|\psi_{SO}\rangle^* \langle\hat{e}_L, -\ell|\psi_{SO}\rangle) \\ S_2^\ell = 2\text{Im}(\langle\hat{e}_R, +\ell|\psi_{SO}\rangle^* \langle\hat{e}_L, -\ell|\psi_{SO}\rangle) \\ S_3^\ell = |\langle\hat{e}_R, +\ell|\psi_{SO}\rangle|^2 - |\langle\hat{e}_L, -\ell|\psi_{SO}\rangle|^2 \end{cases} \quad (3)$$

---

* These authors contribute equally to this work
† E-mail: zhuzhihan@hrbust.edu.cn

Here $\hat{\sigma}_0^{SO}$ and $\hat{\sigma}_{1-3}^{SO}$ are the unitary operator and Pauli matrices for the SOC space, respectively, $S_0^\ell = \sqrt{(S_1^\ell)^2 + (S_2^\ell)^2 + (S_3^\ell)^2} \leq 1$ represents the conservation of probability and $S_0^\ell = 1$ for pure states. From a geometric viewpoint, as shown in Fig. 1, $S_3^\ell$ specifies the latitude of the HOPS corresponding to $a_1^2, a_2^2$; while $S_1^\ell$ and $S_2^\ell$ codetermine the longitude corresponding to the intramode phase $\phi^\ell$ between $|0\rangle$ and $|1\rangle$. That is, $S_{0-3}^\ell$ are a set of complete MU-observables in the two-level SOC space, and one can use them to reconstruct the density matrix of a state through a relation $\hat{\rho}_{SO} = 1/2 \sum_{0-3} S_n^\ell \hat{\sigma}_n$, referred as HOST. Notice that, here one can only determine a concise state $|\psi_{SO}\rangle$ because the knowledge of $A_{|\ell|}(r,\theta,z)$ has been lost in the measurement due to the wavefront-flattening operation.

To measure the full wavefunction of a vector mode, the key resides in the spatially complex probability amplitudes of $\psi_{R,L}^\ell$, more specifically, in acquiring both the intensity profile and associated wavefront, as well as their intramodal phase. For scalar Gaussian photons, a similar technique has been demonstrated using weak measurement [27]. Importantly, for vector photons, the internal SOC structure already provides an observation interface to measure directly $\psi_{R,L}^\ell$. More precisely, if we reduce the observation area of $\hat{\sigma}_{0-3}^{SO}$ into the spin subspace, i.e., a set of MU-observations $\langle\psi_{CV}|\hat{\sigma}_{0-3}^{spin}|\psi_{CV}\rangle$ with respect to polarizations $\hat{e}_{L,R}$, $\hat{e}_{H,V} = (\hat{e}_L \pm \hat{e}_R)/\sqrt{2}$ and $\hat{e}_{D,A} = (\hat{e}_L \pm i\hat{e}_R)/\sqrt{2}$, corresponding observables will become a set of expectation functions in the polar coordinates $\{r,\theta\}$, referred as spatial SPs, which are given by

$$\begin{cases} S_0(r,\theta) = |\langle\hat{e}_R|\psi_{CV}\rangle|^2 + |\langle\hat{e}_L|\psi_{CV}\rangle|^2 = |a_1\psi_R^\ell|^2 + |a_2\psi_L^\ell|^2 \\ S_1(r,\theta) = 2\text{Re}(\langle\hat{e}_R|\psi_{CV}\rangle^* \langle\hat{e}_L|\psi_{CV}\rangle) = |\psi_H^\ell|^2 - |\psi_V^\ell|^2 \\ S_2(r,\theta) = 2\text{Im}(\langle\hat{e}_R|\psi_{CV}\rangle^* \langle\hat{e}_L|\psi_{CV}\rangle) = |\psi_D^\ell|^2 - |\psi_A^\ell|^2 \\ S_3(r,\theta) = |\langle\hat{e}_R|\psi_{CV}\rangle|^2 - |\langle\hat{e}_L|\psi_{CV}\rangle|^2 = |a_1\psi_R^\ell|^2 - |a_2\psi_L^\ell|^2 \end{cases} \quad (4)$$

Here $\psi_{H,V,D,A,L,R}^\ell$ denote polarization-dependent spatial modes of $|\psi_{CV}\rangle$ with respect to the horizontal, vertical, diagonal, antidiagonal, left- and right-circular polarizations, respectively. Moreover, $\psi_{H(D)}^\ell$ and $\psi_{V(A)}^\ell$ are not usually orthogonal to each other only in the case when $|\psi_{CV}\rangle$ is fully entangled, or rather $a_1^2 = a_2^2 = 0.5$ ($\chi^\ell = 0.5\pi$).

It is worth highlighting that $S_{0-3}(r,\theta)$ completely record the spatial complex probability amplitudes of $\psi_{R,L}^\ell$ and their intramode phase. More specifically, $S_0(r,\theta)$ represents the intensity profile of $\psi_{R,L}^\ell$, while $S_{1-3}(r,\theta)$ record the spatial wavefronts of $\psi_{R,L}^\ell$ and their intramode phase $\phi^\ell$ (see Fig. 1). In this way, on the basis of knowing in advance the spin bases of the SOC space, one can reconstruct the realistic vector profile of $\psi_{CV}(r,\theta,z,\ell)$ at the measuring plane $z_0$ through a relation

$$\mathbf{P}(r,\theta) = A_{|\ell|}^2(r,\theta,z_0) \otimes \hat{\rho}_S(r,\theta), \quad (5)$$

where $A_{|\ell|}^2(r,\theta,z_0) = S_0(r,\theta)$ and $\hat{\rho}_S(r,\theta) = 1/2 \sum_{0-3} S_n(r,\theta)\hat{\sigma}_n$ is a functional density matrix describing the *intensity* and *polarization* profiles of measured photons. According to Eq. (4) and (5), for a given full SOC state $\psi_{CV}(r,\theta,z,\ell)$, one can theoretically derive its $S_{0-3}(r,\theta)$ and predict corresponding $\mathbf{P}(r,\theta)$ upon propagation in $z$; while for a given vector mode to be measured at $z_0$, one can observe its $\mathbf{P}(r,\theta)$ in experiment.

Due to the capability of measuring vector profiles of spatial-variant SOPs, the MU-observations $\langle\psi_{CV}|\hat{\sigma}_{0-3}^{spin}|\psi_{CV}\rangle$ are therefore termed spatial-Stokes measurement (SSM) or 'spatial polarimetry' introduced by Fridman *et al.* [33]. However, for the natural question — *How to determine the wavefunction via an observed vector profile?* — it is still an issue that needs to be solved, even in the case of having previous knowledge about the SOC space (HOPS). For this question, if there is one and only one $\hat{\rho}_{SO}$ (or $|\psi_{SO}\rangle$) associated with the observed polarization profile $\hat{\rho}_S(r,\theta)$, one could definitely determine the matching state through a certain 'reverse engineering'. Although this assumption has yet to be proved, it is tenable by giving an appropriate restriction for the SOC space, such as being limited to CV modes.

To determine the SOC state, one need to unearth two salient information from the observed $\hat{\rho}_S(r,\theta)$ are: (i) the topological charges of $\psi_{L/R}^\ell$ for determining HOSP; (ii) and associated $S_{0-3}^\ell$ (or $\{\chi^\ell, \phi^\ell\}$) for locating the position on the sphere. To achieve this, one can find the clue from the morphological feature of observed $\hat{\rho}_S(r,\theta)$ that manifests in functions of spatial polarization ellipticity $\chi(r,\theta)$ and orientation $\phi(r,\theta)$, respectively, which are given by

$$\chi(r,\theta) = \frac{1}{2}\sin^{-1}\left(\frac{S_3(r,\theta)}{S_0(r,\theta)}\right) \text{ and } \phi(r,\theta) = \frac{1}{2}\tan^{-1}\left(\frac{S_2(r,\theta)}{S_1(r,\theta)}\right). \quad (6)$$

First, to obtain the topological charge, consider the fact that the spatial-variant $\hat{\rho}_S(r,\theta)$, or the change of $\phi(r,\theta)$ upon $\theta$, arise from the spatial geometric phase $\pm\ell\theta$ leading to a spatial-variant intramode phase $\phi(r,\theta) = \phi^\ell + 2\ell\theta$ between $\psi^\ell_{L/R}$. Therefore, one can directly acquire the topological charge by interrogating the profile of $\phi(r,\theta)$. Second, to locate the position $\{\chi^\ell, \phi^\ell\}$, after specifying the HOPS, first, one can determine the global orientation $\phi^\ell$ that corresponds to the intramode phase of CV modes through the relation

$$\phi(r,\theta) - 2\ell\theta = \phi^\ell + \gamma(r,\theta). \quad (7)$$

Here $\gamma(r,\theta)$ denotes spatial phase noise and for an ideal pure state $\gamma(r,\theta) \equiv 0$, which can be used to evaluate the 'beam quality' of vector modes. Second, note that $\chi(r,\theta)$ should be uniform over the transverse plane for CV modes. That is one can determine the $\chi^\ell$ and associated $S_3^\ell$ by calculating the mean value of $\chi(r,\theta)$ over the transverse plane, which is given by

$$\chi^\ell = \bar{\chi}(r,\theta)_\perp. \quad (8)$$

Besides, because $\chi^\ell$ specifies the latitude on HOPS, having a similar formulation with the 'concurrence'. Thus, one can determine the degree-of-entanglement (DOE) of measured states through the relation $C(|\psi\rangle) = \sqrt{1 - \bar{\chi}^2(r,\varphi)_\perp}$. About this issue, a related work also reports characterization the DOE of CV modes via SSM recently [34].

*Experiment.* — First of all, it is important to note that, for both HOST and SSM, the measurement apparatus and measured states must be fixed in the same reference frame, and here we chose the tabletop as the horizontal plane. Figure 1 shows the experimental procedure, where we used an attenuated coherent light (780 nm), collimated from a single-mode fiber and passing through a polarizing beam splitter (PBS), to generate TEM$_{00}$-mode photons with horizontal polarization. The H-polarized photons are then converted into desired CV modes by passing through a pair of waveplates and a q-plate (*Thorlabs* WPV10L). The prepared CV modes are then divided equally by a non-polarizing beam splitter (NPBS), and at the reflected side, they are measured by a HOST apparatus. While at the transmission side, a set of waveplates together with a PBS are employed to perform the spatial Stokes tomography, and a CCD (PI-MAX4) is used to record polarization-dependent spatial modes. In experiments, by adjusting the angles of waveplates in front of the q-plate, we first prepared a group of rotational-invariant CV modes ($\ell = +1$) as the targets to be measured, whose positional trajectories on the sphere correspond to the red paths and associated (equally spaced) red dots, as shown in Fig. 1(a). We then measured the prepared modes via SSM and HOST, simultaneously, and compared the results obtained via the two approaches.

*Results.* — Figure 2(a) shows the observed vector profiles of the target modes, where their polarization profiles are typically rotational-invariant CV modes with a few defects (i.e., not perfectly pure states for an observer on the HOPS) and the intensity profiles are feature of Hypergeometric-Gaussian (HyGG) modes. Figure 2(b) shows their spatial-polarization orientation (or geometric-phase profiles) $\phi(r,\theta)$ and associated global intramode phase with spatial-phase noise $\phi^\ell + \gamma(r,\theta)$, respectively. Here we see that, first, the profiles of $\phi(r,\theta)$ indicate their topological charge $\ell = 1$; and second, the slices of $\phi^\ell + \gamma(r,\theta)$ are not flat, giving rise to the 'defects' in their vector profiles. On the basis of above data, Fig. 2(c) shows their density matrices of determined $|\psi_{SO}\rangle$ by using Eq. (7) and (8). For intuitive illustration and comparison, Fig. 2(d) shows the positions of SOC states on the HOPS determined by SSM (blue points) and HOST (orange points), respectively, as well as their HOSPs and concurrence chart.

*Analysis.* — From Fig. 2(d) we note that all the states determined via SSM (blue) and HOST (orange) deviate slightly from each other, which manifests in states on HOPS, associated $S_{0-3}^\ell$, and the concurrence. Besides, the error bars in SSM data are several orders of magnitude larger than those in HOST. To analyze the reason for these differences, first of all, notice that the SOC conversion efficiency of q-plates (*Thorlabs* WPV10L) we used for preparing CV modes and performing HOST is approx. 99%. That is the prepared CV modes are not completely pure states for an observer on the HOPS. As a result, in theory, four real-valued $S_{0-3}^\ell$ observed via HOST cannot provide exact and enough knowledge of the vector profile of measured photons. While, due to no wavefront-flattening operations, SSM can completely record the realistic SOC structure of measured photons. Namely, the bigger error bars in SSM data depict the quality and purity of determined states on a specific HOPS. From this point, SSM shall provide a more accurate SOC state determination and can be extended to apply to more general states (even mixed states). Note, we have assumed all states to be measured are pure in the calculation. Additionally, compared with HOST approach, SSM need much more photons to record the spatial modes completely and precisely, and we will further discuss this aspect in a separate paper.

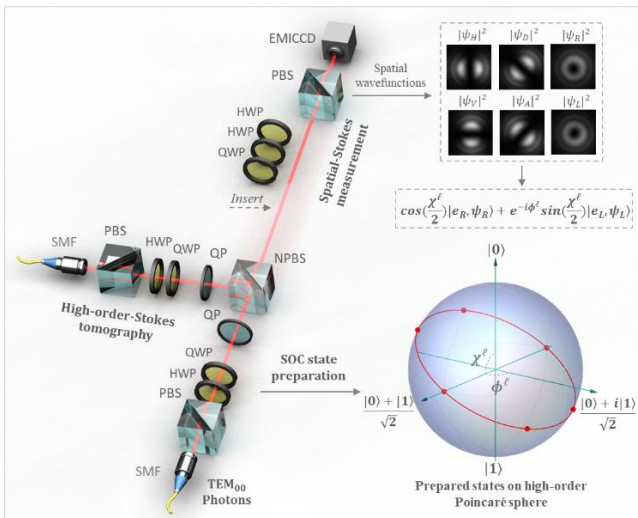

FIG. 1. Experimental setup. The abbreviations of key components include the non-polarizing beam splitter (NPBS), polarizing beam splitter (PBS), half-wave plate (HWP), quarter-wave plate (QWP) and q-plate (QP), single-mode fiber (SMF).

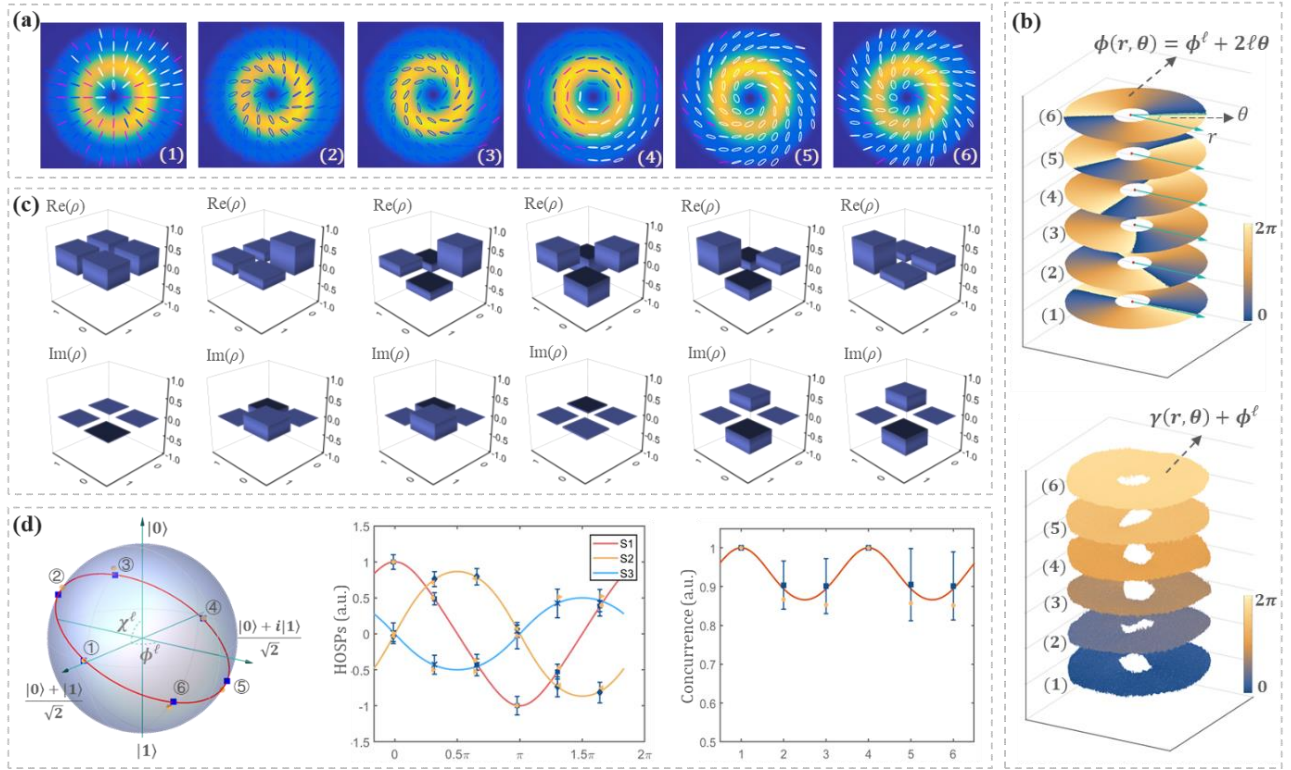

FIG. 2. Experimental results. (a) Observed vector profiles, where blue, red and white ellipse fields represent the right-hand, linear, and left-hand chirality of polarizations, respectively. (b) Observed geometric-phase profiles and associated global intramode phase with spatial-phase noise. (c) Density matrices of determined SOC states via SSM, where $|0\rangle = |\hat{e}_R, +\ell\rangle$ and $|1\rangle = |\hat{e}_L, -\ell\rangle$. (d) Measured states on the HOPS, associated HOSPs, and calculated concurrence, where orange and blue markers are experimental results determined via HOST and SST, respectively.

To demonstrate the validity of states determination, one can compare the theoretical vector profile of a determined $|\psi_{SO}\rangle$ and corresponding experimental observation. More specifically, by using Eq. (4) and (5), we first simulate the vector profiles of SOC states determined via HOST, SSM, and SSM with the correction of q-plate efficiency (99%), as shown in Fig. 3(a)-(c), respectively. In the simulation, for states determined via HOST, consider the knowledge of $A_{|\ell|}(r,\theta,z)$ has been lost in the measurement, for this, we choose propagation constant Laguerre-Gaussian (LG) modes as the spatial mode; while for states determined via SSM, we employ HyGG modes with $z_R = 0.15$ as the spatial mode (*see Appendix A*). [35,36]

Here we see that the corrected vector profiles shown in Fig. 3(c) are well in accordance with the experimental observation shown in Fig 2(a). For the sake of clarity, we also compare the change of $S_{1-3}(r,\theta)$ as function of $\theta$ for the observed polarization profile and theoretical profiles determined via HOST, SSM, and corrected SSM, respectively, as shown in Fig. 3(d). We see that the data of SSM determination agree with the observation better than those determined through HOST, while the data of corrected SSM determination matches the best with the observation. We then investigated the fidelity of $\hat{\rho}_S(r,\theta)$ between determined states and experimental observations, or rather the 'spatial-variant SOP autocorrelation', which can be expressed as $F_{SOC} = \langle \hat{\rho}_S(r,\theta)_{Theor.} | \hat{\rho}_S(r,\theta)_{Obser.} \rangle^2$. Fig. 3(e) shows the calculation, we see that, in general, the fidelity of SSM determination is better than HOST, which is especially so for the corrected case. Besides, we note that the error bars of SSM are smaller than HOST and the corrected SST is smallest. Note, the larger error bars of SSM data shown in Fig. 2(d) indicate the measured states are not pure for the predefined HOSP; while the smaller error bars in the fidelity shown in Fig. Fig. 3(e) manifest the determined states are closer to the realistic SOC structures of measured photons.

In addition to the above rotational-invariant CV modes, this approach can also apply to more general CV modes, such as anti-vortex modes for $\ell = -1$ and high-order CV modes for $|\ell| > 1$, and even Full-Poincare modes (*see Appendix B*). Yet considering their vector profiles are no longer rotationally invariant, one has to maintain the measuring apparatus and measured states in the same reference frame. It should be pointed out, however, that the general CV modes are essential resources for high-dimension quantum information. Moreover, in a practical application, it is inevitable to avoid the relative rotation of reference frame, such as misalignment between systems or birefringence transmission. Namely, it is of greater significance to determine the SOC state of a more general CV mode that has experienced a rotation of the reference frame during propagation. This task is impossible for the standard QST based on HOST, because the observed probabilities in this case cannot provide valid information to estimate the wavefunction. While for the SSM approach, one can still observe a vector profile of measured photons in a 'rotated reference frame', and based on this work, we will further demonstrate how to determine the original SOC state from this reference-frame rotated vector profile in a future paper.

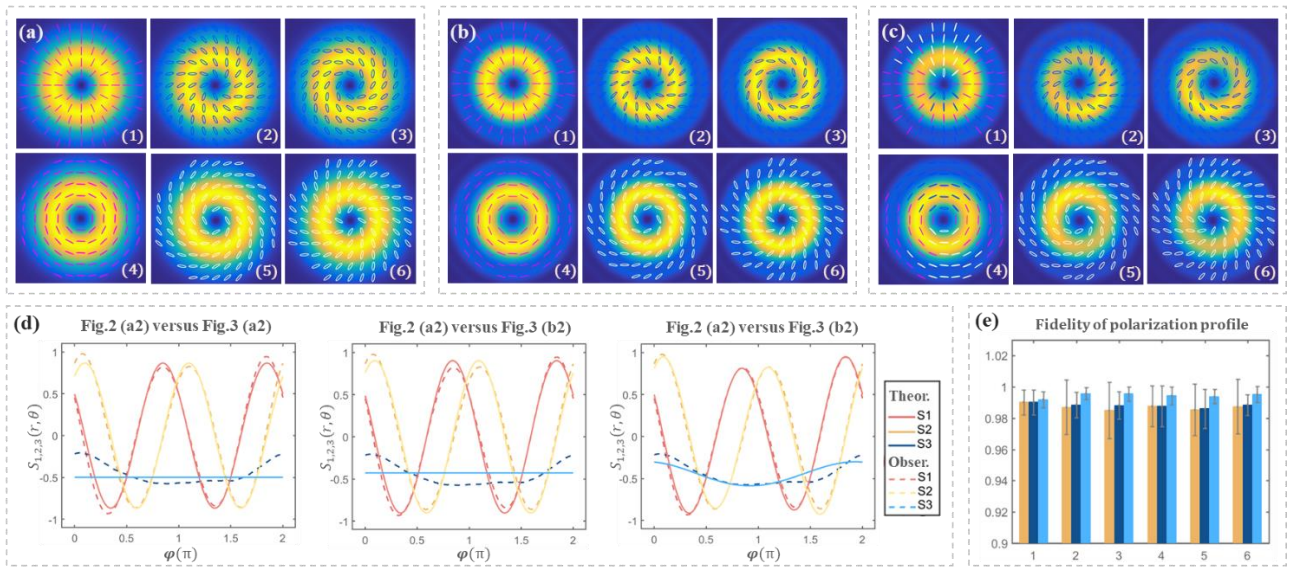

FIG. 3. (a)-(c) Simulated vector profiles of SOC states determined via HOST (*pure state*), SSM (*pure state*), and corrected SSM (*mixed state*), respectively. (d) The comparison of $S_{1-3}(r,\theta)$ upon $\theta$ for determined states and experimental observation. (e) Calculated polarization-profile Fidelity of SOC states determined via HOST (*yellow*), SSM (*deep blue*) and corrected SSM (*light blue*), respectively.

*Conclusion.* — In summary, we have demonstrated a direct determination of SOC state via SSM. Compared with the standard QST approach based on HOST, our results show that the SSM approach can not only acquire a realistic vector profile of measured photons but can also determine the wavefunction in a more accurate and precise way. This advantage makes it ideal for in-situ detecting SOC states in communication channels and characterizing the quality of photonic SOC sources or associated geometric-phase devices. Besides, this approach has the potential to be extended to more general SOC states, even in the case when observers have lost the knowledge about the shared reference frame, which is crucial in the practice of high-dimension quantum information.

**References**

1. K. Y. Bliokh, F. J. Rodriguez-Fortuno, F. Nori, *et al.* Spin-orbit interactions of light. Nat. Photon. **9**, 796, (2015).
2. A. Aiello, F. Töppel, C. Marquardt, E. Giacobino, and G. Leuchs. Quantum-like nonseparable structures in optical beams. New J. Phys. **17**, 043024 (2015).
3. C. V. S. Borges, M. Hor-Meyll, J. A. O. Huguenin, and A. Z. Khoury. Bell-like inequality for the spin-orbit separability of a laser beam. Phys. Rev. A **82**, 033833 (2010).
4. J. Chen, C.-H Wan, and Q.-W. Zhan. Vectorial optical fields: recent advances and future prospects. Sci. Bull. **63**, 54 (2018).
5. Zhan Q. Cylindrical vector beams: from mathematical concepts to applications. Adv. Opt. Photon. **1**,1 (2009).
6. H. Rubinsztein-Dunlop, *et al.* Roadmap on structured light. J. Opt. **19**, 013001 (2017).
7. C. Rosales-Guzmán, B. Ndagano and A. Forbes. A review of complex vector light fields and their applications. J. Opt. **20**, 123001 (2018).
8. R. Chen, K. Agarwal, C. J. R. Sheppard, and X. Chen, Imaging using cylindrical vector beams in a high-numerical-aperture microscopy system. Optics lett. **38**, 3111-4 (2013).
9. L. Du, A. Yang, A. V. Zayats, X. Yuan. Deep-subwavelength features of photonic skyrmions in a confined electromagnetic field with orbital angular momentum. Nat. Phys. 2019: 1.
10. V. D'Ambrosio, et al. Complete experimental toolbox for alignment-free quantum communication. Nat. Commun. **3**:961 (2012).
11. G. Vallone, et al. Free-Space Quantum Key Distribution by Rotation-Invariant Twisted Photons, Phys. Rev. Lett. 113, 060503 (2014).
12. V. D'Ambrosio, et al. Entangled vector vortex beams. Phys. Rev. A **94**, 030304(R) (2016).
13. S. Berg-Johansen, F. Töppel, B. Stiller, P. Banzer, M. Ornigotti, E. Giacobino, G. Leuchs, A. Aiello, and C. Marquardt, Classically entangled optical beams for high-speed kinematic sensing. Optica **2**, 864-868 (2015).
14. D. Naidoo, et al., Controlled generation of higher-order Poincare sphere beams from a laser. Nat. Photon. **10**, 327-332 (2016).
15. S. Slussarenko, A. Alberucci, C. P. Jisha, B. Piccirillo, E. Santamato, G. Assanto, L. Marrucci. Guiding light via geometric phases. Nat. Photon. **10**, 571-575 (2016).
16. F. Cardano, L. Marrucci. Spin-orbit photonics. Nat. Photon. **9**, 776 (2015).
17. P. Chen, et al. Digitalizing Self-Assembled Chiral Superstructures for Optical Vortex Processing. Adv. Mater. **30**(10), 1705865 (2018)
18. J. Vieira, R. M. G. M. Trines, E. P. Alves, R. A. Fonseca, J. T. Mendonça, R. Bingham, P. Norreys, and L. O. Silva. Amplification and generation of ultra-intense twisted laser pulses via stimulated Raman scattering. Nat. Commun. **7**, 10371 (2016).
19. H.-J. Wu, Z.-Y. Zhou, W. Gao, B.-S. Shi, and Z.-H. Zhu. Dynamic tomography of the spin-orbit coupling in nonlinear optics. Phys. Rev. A **99**, 023830 (2019).
20. G. Milione, H. I. Sztul, D. A. Nolan, and R. R. Alfano, Higher-Order Poincaré Sphere, Stokes Parameters, and the Angular Momentum of Light. Phys. Rev. Lett. **107**, 053601 (2011).

21. G. Milione, S. Evans, D. A. Nolan, and R. R. Alfano, Higher Order Pancharatnam-Berry Phase and the Angular Momentum of Light. Phys. Rev. Lett. **108**, 190401 (2012).
22. L. J. Pereira, A. Z. Khoury, K. Dechoum. Quantum and classical separability of spin–orbit laser modes. Phys. Rev. A **90**, 053842 (2014).
23. M. McLaren, T. Konrad, and A. Forbes, Measuring the non-separability of vector vortex beams Phys. Rev. A **92**, 023833 (2015).
24. B. Ndagano, H. Sroor, M. McLaren, C. R. Guzmán, and A. Forbes. A beam quality measure for vector beams. Opt. Lett. **41**, 3407-3410 (2016).
25. B. Pinheiro da Silva, D. S. Tasca, E. F. Galvão, and A. Z. Khoury. Astigmatic tomography of orbital-angular-momentum superpositions. Phys. Rev. A **99**, 043820 (2019).
26. H. Qassim, F. M. Miatto, J. P. Torres, M. J. Padgett, E. Karimi, and R. W. Boyd. Limitations to the determination of a Laguerre–Gauss spectrum via projective, phase-flattening measurement. J. Opt. Soc. Am. B **31**, A20 (2014).
27. Lundeen, J. S., Sutherland, B., Patel, A., Stewart, C. & Bamber, C. Nature 474, 188–191 (2011).
28. Jeff Z. Salvail, M. Agnew, A. S. Johnson, E. Bolduc, J. Leach, & R. W. Boyd. Full characterization of polarization states of light via direct measurement. Nat. Photon. **7**, 316–321 (2013).
29. Malik, M. et al. Direct measurement of a 27-dimensional orbital-angular-momentum state vector. Nat. Commun. **5**: 3115 (2014).
30. J. S. Lundeen and C. Bamber. Procedure for Direct Measurement of General Quantum States Using Weak Measurement. Phys. Rev. Lett. **108**, 070402 (2012).
31. G. S. Thekkadath, L. Giner, Y. Chalich, M. J. Horton, J. Banker, and J. S. Lundeen. Direct Measurement of the Density Matrix of a Quantum System. Phys. Rev. Lett. **117**, 120401 (2016).
32. Kim, Y. et al. Direct quantum process tomography via measuring sequential weak values of incompatible observables. Nat. Commun. **9**, 192 (2018).
33. M. Fridman, M. Nixon, E. Grinvald, N. Davidson, and A. Friesem, Real-time measurement of space-variant polarizations, Opt. Express **18**, 10805-10812 (2010).
34. Adam Selyem et al. Basis independent tomography of complex vectorial light fields by Stokes projections. arXiv: 1902.07988v1
35. Z.-Y. Zhou, Z.-H. Zhu, S.-L. Liu, Y.-H Li, S. Shi, D.-S. Ding, L.-X. Chen, W. Gao, G.-C. Guo, and B.-S. Shi, Quantum twisted double-slits experiments: confirming wavefunctions' physical reality. Sci. Bull. **62**,1185 (2017).
36. E. Karimi, G. Zito, B. Piccirillo, L. Marrucci, and E. Santamato. Hypergeometric-Gaussian modes. Opt. Lett. **32**, 3053 (2007).